\documentclass[letter]{aa}

\usepackage{hyperref}
\usepackage{placeins}

\hypersetup{
        colorlinks = true,
        linkcolor = {red},
        citecolor = {blue},
        linkbordercolor = {white},
        citebordercolor = {purple}
}
\usepackage{graphicx}
\usepackage{txfonts}
\usepackage{capt-of}

\begin{document}

   \title{Post-DART impact ejecta re-accumulation on Dimorphos and implications for Hera gravity measurements }

\author{S. D. Raducan\inst{1,2}
        \and
        \"O. Karatekin\inst{2}
        \and
        M. Popescu\inst{3,4}
        }
\institute{Archaeology, Environmental Changes \& Geochemistry (AMGC), Vrije Universiteit Brussel, Brussels, Belgium
      \and
           Reference Systems and Planetology Department, Royal Observatory of Belgium, Brussels, Belgium
      \and
          Institute of Space Science – INFLPR subsidiary, 409, Atomi\c{s}tilor Street, 077125 M\u{a}gurele, Ilfov, Romania
      \and
          University of Craiova, Str. A. I. Cuza nr. 13, 200585 Craiova, Romania
      }

   \date{ }

  \abstract

\abstract
  {The DART impact might have substantially modified the surface and near-surface structure of Dimorphos, affecting the morphology and gravity field that is to be measured by ESA's Hera mission.}
  {We investigate the post-impact redistribution of ejecta and its dependence on the assumed internal structure of Dimorphos.}
  {We performed high-resolution SPH simulations of the DART impact into two low-cohesion rubble-pile models, with and without a large central boulder. We followed the processes of ejecta re-accumulation and gravity evolution for six hours.}
  {The impact ejects $\approx13\%$ of Dimorphos's mass, most of which is re-accreted, producing near-global resurfacing. The ejecta blanket exceeds 8~m around the impact depression but is only a few tens of centimetres thick near the antipode, with similar large-scale deposition in both interior models. Local gravity changes reach about 10\% of the pre-impact value. The central boulder affects the surface gravity only within the impact depression, where the removal of overlying material enhances its gravitational contribution.}
  {Hera will observe a surface and gravity field substantially modified by DART, which must be considered when interpreting measurements in terms of the internal structure of Dimorphos.}


   \maketitle


\section{Introduction}

On 26 September 2022, NASA's Double Asteroid Redirection Test (DART) spacecraft impacted Dimorphos, the secondary of the Didymos binary asteroid system, and successfully altered its orbit around Didymos \citep{Daly2023, Cheng2023, Chabot2024}. The impact produced an amount of ejecta, which was observed by ground- and space-based telescopes \citep{Li2023} and by the LICIACube \citep{Dotto2024}.
The measured momentum transfer and morphology of the evolving ejecta suggest that Dimorphos is a low-cohesion rubble-pile asteroid \citep{Raducan2024}. However, different interior structures can produce outcomes that are consistent with existing observations. These differences might be reflected in the distribution of re-accumulated ejecta and the post-impact gravity field, providing diagnostics of the surface and internal properties of Dimorphos.

ESA's Hera mission \citep{Michel2022} will arrive at the Didymos system in late 2026 to complete the DART experiment. Hera will determine the shape, mass, composition, and dynamical properties of the binary, while the Juventas CubeSat will probe Dimorphos with the JuRa low-frequency radar and measure the local gravity vector with the GRASS gravimeter after landing \citep{Michel2022}. Together with hyperspectral observations from HyperScout-H \citep{Popescu2025,Prodan2026}, these measurements will constrain the surface, near-surface, and interior properties of Dimorphos.

Interpreting Hera observations requires accounting for DART-induced modifications. Numerical simulations \citep[e.g.][]{Raducan2022d,Raducan2024,Raducan2024b} show that much of the excavated material remains bound and is redistributed across the surface, modifying the topography, near-surface mass distribution, and gravity field. To support the interpretation of Hera observations, we investigate how different interior structures affect ejecta evolution and re-accumulation.

   \begin{figure*}[h!]
   \centering
    \includegraphics[width=0.7\linewidth]{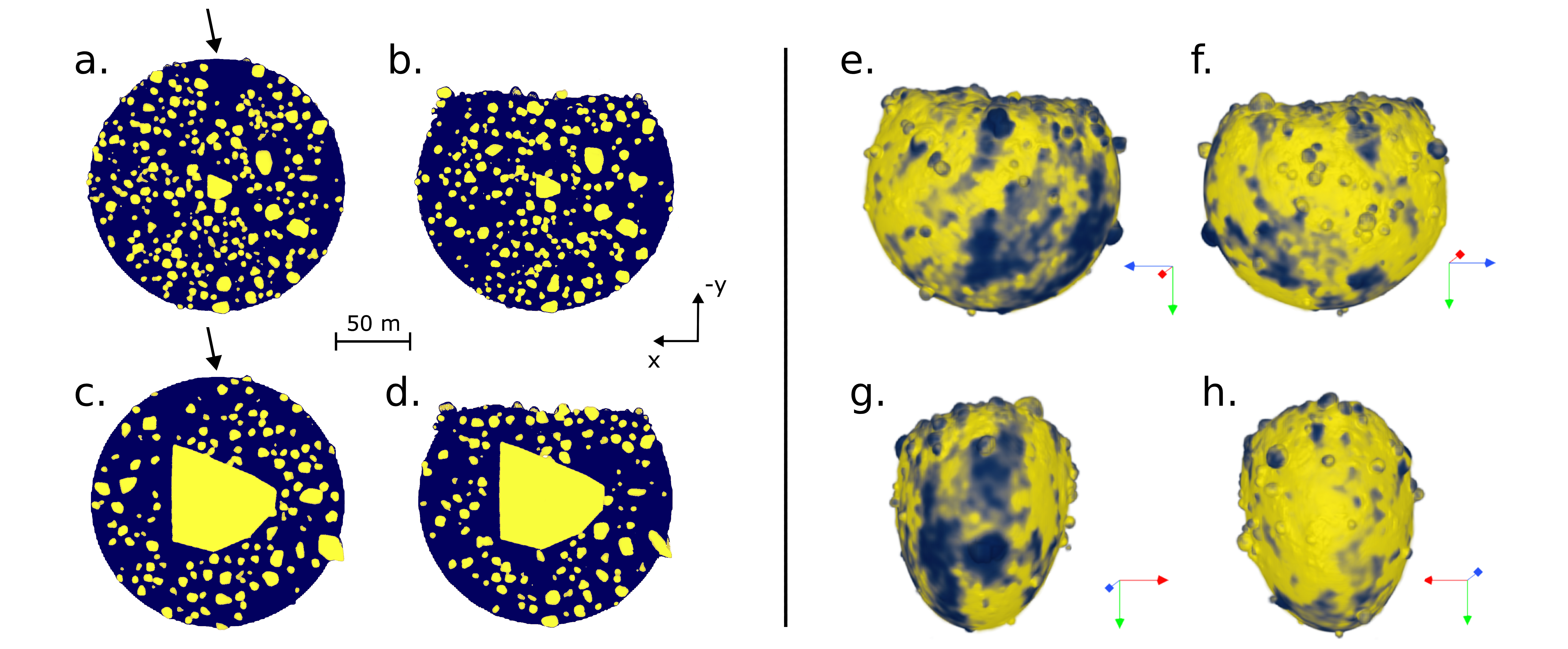}
      \caption{2D target cross-sections and three-dimensional maps of ejecta re-accumulation 6~h after the DART impact. Panels (a) and (b): Homogeneous rubble-pile model before and after impact. Panels (c) and (d): Corresponding central-boulder model. The black arrows in (a) and (c) indicate the impact direction. Panels (e)–(h): Ejecta blanket for the homogeneous rubble-pile model, viewed from the north and south poles, the anti-Didymos hemisphere, and the Didymos-facing hemisphere, respectively. Yellow marks re-accumulated ejecta and blue regions deposition thicknesses $<20$~cm. The ejecta blanket is highly heterogeneous, with large surface boulders locally shielding the underlying terrain.}
         \label{DART_outcome}
   \end{figure*}

\vspace{-0.2cm}
\section{Numerical methods}
\label{sec:methods}

We used the Bern smoothed particle hydrodynamics (SPH) code \citep{Benz1999,Jutzi2008} to model the DART impact, boulder fragmentation, ejecta excavation, and re-accumulation. The simulations follow the modelling approach of \citet{Raducan2024}, extended here to investigate different interior structures (Fig.~\ref{DART_outcome}).

The DART spacecraft was represented by a 580~kg aluminium sphere impacting Dimorphos at $6.14$~km~s$^{-1}$, with the impact geometry reconstructed from the DART trajectory \citep{Daly2023}. The projectile was modelled using the Tillotson equation of state (EOS) and a von Mises strength model.
Both the boulders and granular matrix were described using the Tillotson EOS for SiO$_2$ \citep{Tillotson1962,Melosh2007}. The matrix was modelled as a frictional granular material using the Drucker-Prager yield criterion with an internal friction coefficient of $f=0.6$. Its initial porosity was 45\%, corresponding to a bulk density of $\rho_{\rm m}$=1.8~g~cm$^{-3}$, and compaction was treated using the $P$--$\alpha$ model \citep{Jutzi2008}. The boulders were described by a tensile damage and fragmentation model \citep{Jutzi2015}, with $Y_{\rm T}\approx10$~MPa and an initial microporosity of 10\% ($\rho_{\rm b}=2.9$~g~cm$^{-3}$) (Table~\ref{table:model_parameters}).

To follow the late-stage gravitational evolution, we used the approach of \citet{Raducan2022}, which allows for larger time steps after the shock and fragmentation phases. Each simulation used $\approx8\times10^6$ SPH particles and was continued for 6~h, about half the orbital period of Dimorphos around Didymos \citep{Cheng2023}. By then, 94\% of the ejecta had re-accreted or escaped; the remaining slowly orbiting material, which may evolve over days to weeks, was not followed further (Appendix~\ref{app:re-accumulation}).

We considered two end-member models for Dimorphos. Both reproduce the pre-impact shape, approximated as a triaxial ellipsoid with principal axes $\approx177 \times 151 \times 116$~m \citep{Daly2023}. The first model consists of non-spherical boulders embedded within a granular matrix, with a differential size-frequency distribution of slope $-3.4$, consistent with the pre-DART surface of Dimorphos \citep{Pajola2024}. The resolved boulders range from 2.2 to 18~m  (Fig.~\ref{DART_outcome}a). 
The second model has the same rubble-pile structure but contains an additional 64~m boulder at its centre (Fig.~\ref{DART_outcome}c).
The resulting target masses are $3.53\times10^9$~kg and $3.79\times10^9$~kg respectively. Both are lower than the nominal mass from the reconstructed shape and assuming equal bulk densities for Didymos and Dimorphos \citep{Daly2023}, but are closer to the more recent dynamical estimate of \citet{Makadia2026} (M$_{\rm{D}} \sim2.7 \times 10^9$ kg).

The gravitational acceleration was computed directly from the evolving SPH particle distribution, accounting for changes in Dimorphos' shape and internal mass distribution. Didymos was represented as a point mass ($M_{\rm Didy}=5.28\times10^{11}$~kg) at a separation of 1.2~km \citep{Cheng2023}. The simulations were performed in the co-rotating frame, including gravitational, centrifugal, and Coriolis accelerations. 
At $t=6$~h, the surface-gravity field was calculated from the mass distribution given by the SPH particle positions and masses  (Appendix~\ref{app:gravity}). To isolate internal density heterogeneities from shape effects, we also calculated a normalised Bouguer anomaly relative to a uniform-density body (Appendix~\ref{app:bouguer}).

\section{Results}

\subsection{Post-impact target morphology}

We first examined the morphology of Dimorphos following the DART impact. For
both interior configurations, the impact deforms the target, consistent with
previous results for low-cohesion rubble-pile models of Dimorphos
\citep{Raducan2024, Sanchez2025}. Approximately 13\% of the target mass is launched as ejecta (Appendix~\ref{app:re-accumulation}), while only $\approx 1\%$ escapes Dimorphos, and the rest re-accumulates onto its surface.
Rather than a bowl-shaped crater, the excavation flow generates a
flat-floored depression accompanied by large-scale rearrangement of
surrounding material (Fig.~\ref{DART_outcome}). The morphology at $t=6$~h is
similar for the two models (Fig.~\ref{DART_outcome}b--d). The central boulder
lies below the excavation zone, at depths where the shock has decayed
sufficiently that the impedance contrast has little effect on excavation  \citep{Raducan2020}.

   \begin{figure}[t]
   \centering
   \includegraphics[width=0.89\linewidth]{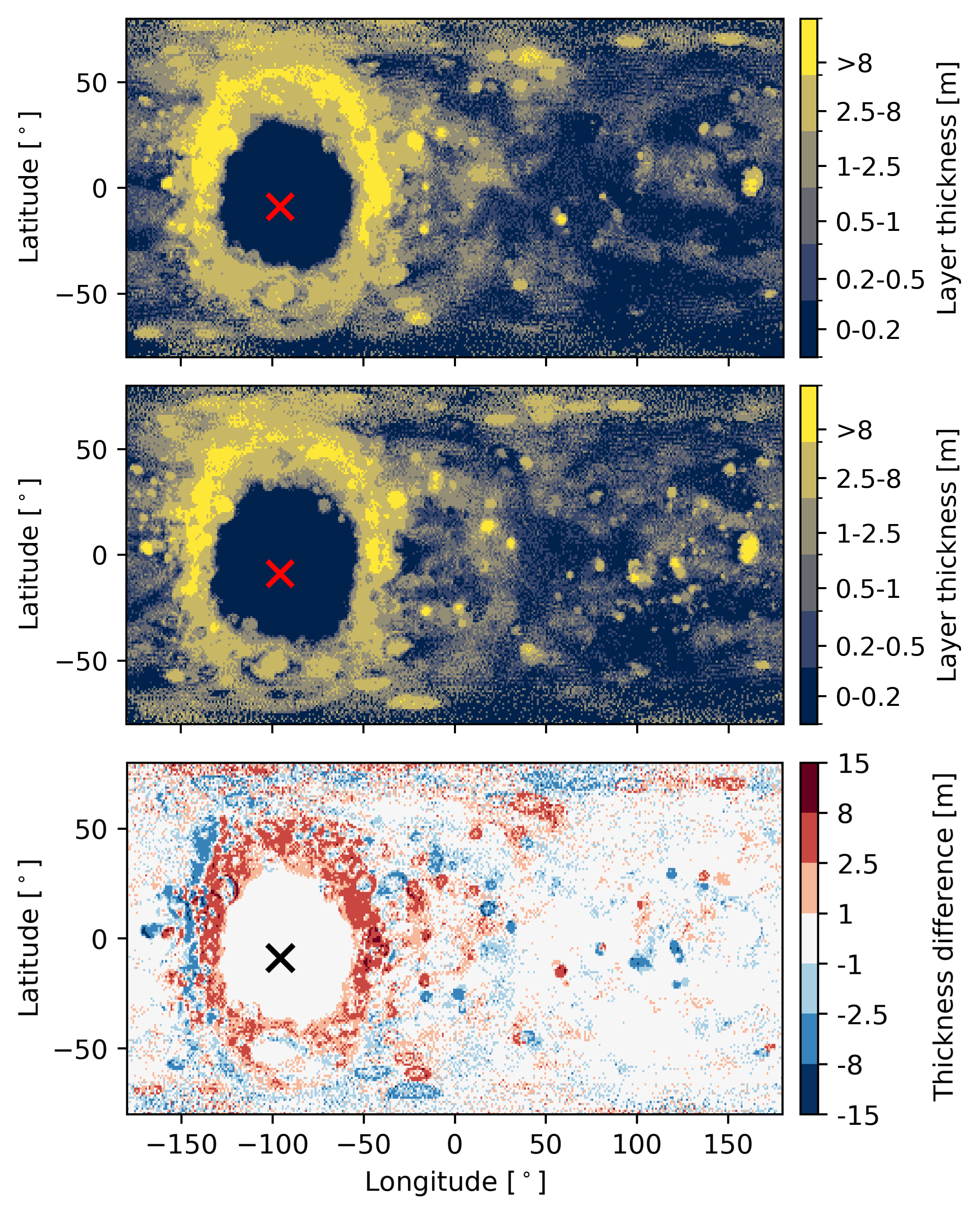}
      \caption{Maps of the re-accumulated ejecta layer thickness 6~h after the DART impact for the homogeneous rubble-pile model (top) and the central-boulder model (middle). Bottom: Thickness difference between the two models (top minus middle).      }
         \label{ejecta_map}
   \end{figure}

   \begin{figure*}
   \centering
   \includegraphics[width=0.89\linewidth]{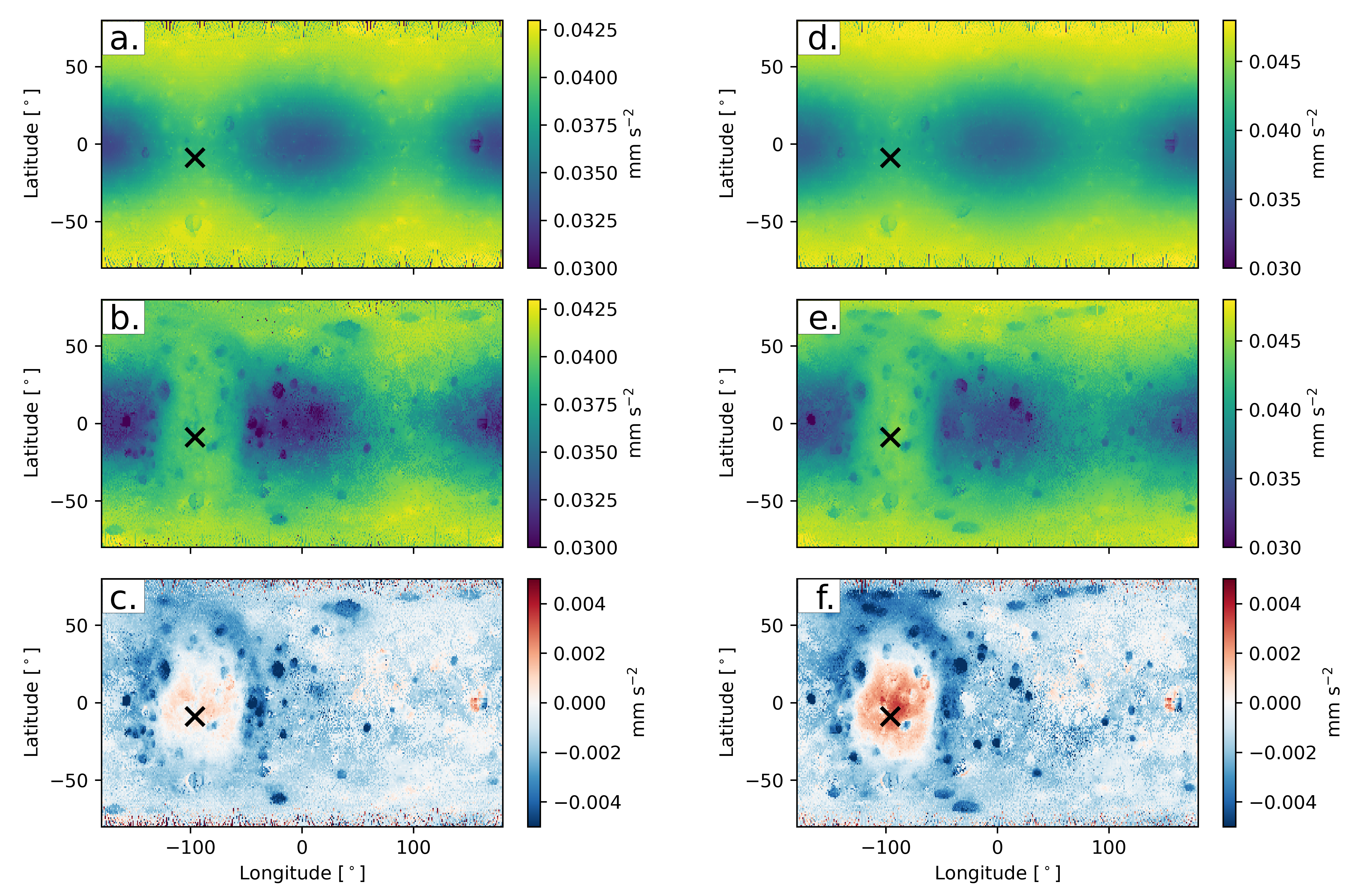}
   \caption{Maps of the effective surface gravity (in mm s$^{-2}$) before and 6~h after the DART impact. Panels (a) and (b): Homogeneous rubble-pile model. Panels (d) and (e): Central-boulder model. Panels (c) and (f): Corresponding differences between the post-impact and pre-impact gravity maps. The black cross marks the impact location.}
    \label{gravity}
    \end{figure*}

\subsection{Ejecta re-accumulation}

To investigate the post-impact resurfacing of Dimorphos, we mapped the thickness of the re-accumulated ejecta layer by binning all particles settled on the surface into $1^\circ \times 1^\circ$ longitude--latitude cells and converting the deposited mass in each cell into an equivalent layer thickness (Appendix~\ref{app:layer}). By this time, $\approx$94\% of the ejecta has either re-accreted or escaped the Didymos system, allowing the final resurfacing pattern to be characterised.

Although the entire surface of Dimorphos is covered by re-accreted ejecta, the deposit is highly asymmetric (Fig.~\ref{DART_outcome}). The spatial distribution is strongly influenced by the low-gravity binary environment, where the gravity of Didymos and rotating-frame accelerations modify the ballistic trajectories of slow-moving ejecta, producing regional variations in blanket thickness (Fig.~\ref{ejecta_map}). In comparison, simulations without these effects produce a homogeneous ejecta blanket \citep{Raducan2024}.

The thickest deposits occur around the rim of the impact depression, where the ejecta layer locally exceeds 8~m. In contrast, at the antipodal region (longitude $\sim100^\circ$ at low latitudes) the ejecta layer is only a few tens of centimetres thick (Fig.~\ref{ejecta_map}). 

The SPH simulations indicate that internal heterogeneity influences the ejecta blanket morphology. Large boulders can penetrate the outward-moving ejecta curtain and remain exposed after re-accretion (Fig.~\ref{ejecta_map}), while also shielding nearby surfaces from fresh deposition (Fig.~\ref{DART_outcome}). After landing, both ejecta and boulders may undergo downslope motion driven by local slopes, so the final map reflects the combined effects of ballistic deposition and subsequent surface transport.
Comparing the thickness maps for models with and without a central boulder (Fig.~\ref{ejecta_map}) reveals no systematic differences in the large-scale distribution. Away from the depression rim and individual boulders, they are generally below 1 m. The principal differences are confined to the depression and the final locations of boulders, arising from small changes in the local gravity field produced by the higher-density central boulder. Boulder emplacement therefore depends on the internal configuration, whereas the large-scale resurfacing pattern is largely independent of the assumed interior.

\subsection{Post-impact gravity field}

The DART impact modifies the surface gravity field through global shape changes, local topographic variations, and the redistribution of near-surface material. The shape change primarily reflects material displaced from the impact region toward the surrounding surface.
Figure~\ref{gravity} shows the total surface gravitational acceleration before and at $t=6$~h after the impact for the homogeneous rubble-pile and central-boulder models. Prior to the impact, the gravity field is primarily controlled by the global shape of the asteroid, with small local perturbations produced by shallow boulders. After the impact, the large-scale pattern remains largely unchanged, while the most significant differences are local, where shape changes and material redistribution modify the near-surface mass distribution.

The increase in local gravity anomalies is primarily associated with the
redistribution of metre-scale boulders. Large boulders excavated from the
impact cavity are re-deposited across the surface during ejecta
re-accretion, while impact-induced deformation and large-scale granular
motion expose additional buried boulders. Previous simulations have shown
that metre-scale boulders are preferentially ejected at lower velocities
than the surrounding granular material \citep{Raducan2024}, increasing
their probability of remaining gravitationally bound to Dimorphos.
Consequently, a larger number of boulders remain exposed at the surface,
producing more pronounced local gravity anomalies than before the impact.
Around the impact site, the gravitational acceleration increases as a
result of the combined shape change and redistribution of material. The
difference maps show that these local changes reach about
$0.004$~mm~s$^{-2}$, corresponding locally to $\approx$10\% of the
pre-impact surface gravity.

Despite their different internal structures, the homogeneous rubble-pile
and central-boulder models exhibit similar spatial patterns in the
post-impact gravity field, in both cases dominated by the redistribution of near-surface material. The principal difference is the absolute magnitude of the acceleration, which is systematically higher for the central-boulder model owing to its greater total mass (3.79 vs 3.53 $\times 10^9$~kg).
Away from the impact site, the dense central boulder does not introduce a distinctive spatial signature. The exception is the impact depression, where excavation brings the surface $\approx$20~m closer to the central boulder. Relative to a uniform-density body of identical shape, this produces a local anomaly of $\approx$10\% ($\approx$0.004~mm~s$^{-2}$) that is absent from the homogeneous model (Appendix~\ref{app:bouguer}). This also accounts for the stronger gravity increase within the depression of the central-boulder model (Fig.~\ref{gravity}f compared with Fig.~\ref{gravity}c).

\section{Discussion}

The simulations presented here provide forward predictions for several of the key observations Hera will acquire at Dimorphos. The DART impact not only modified the shape of the asteroid, but also redistributed material over its entire surface and altered its near-surface gravity field. These changes should therefore be considered when interpreting the measurements returned by the Hera spacecraft and its Juventas CubeSat.

The HyperScout-H instrument onboard Hera will acquire hyperspectral images over the $\approx0.650$--$0.950~\mu\mathrm{m}$ range, mapping the surface composition, space-weathering effects, possible exogenous material, and material excavated by the DART impact \citep{Popescu2025}.
Space weathering has been invoked to explain spectral differences between weathered S-type and fresher, ordinary chondrite-like Q-type asteroids. Laboratory experiments show that weathering causes spectral reddening and darkening and reduces absorption-band depths, although spectral properties also depend on composition and surface texture \citep{Brunetto2015}.
Because reflectance spectra sample only the uppermost few microns, even the thinnest predicted deposits exceed this depth, and the near-global ejecta coverage may make most of Dimorphos appear spectrally resurfaced. Remnants of the pre-DART surface are most likely where the least ejecta was
deposited, particularly near the antipode and around large boulders that shielded the surface during emplacement (Fig.~\ref{DART_outcome}).

The GRASS gravimeter will measure the local gravity vector after landing. Our results imply that this measurement might be strongly influenced by the local post-impact environment, particularly by nearby exposed, re-deposited, or shallowly buried boulders and by variations in ejecta thickness. Such local heterogeneities can modify both the magnitude and direction of the gravity vector. The predicted post-impact changes reach about 0.004~mm~s$^{-2}$, which is about two orders of magnitude above the sensitivity of GRASS \citep{Noeker2025}. The sensitivity to the deep interior is greatest within the impact depression, where excavation has removed $\approx$20~m of overlying material and a dense central body would increase the local gravity by $\approx$10\%, comparable to the effect of shallow boulders. However, a measurement at a single location may not be representative of the asteroid as a whole, and near-surface contributions cannot be uniquely separated from deeper mass heterogeneities using gravity alone.

It is is therefore essential to use JuRa low-frequency radar in complement to GRASS. While GRASS is dominated by the shallow mass distribution, JuRa directly probes the interior and can reveal whether the boulders influencing the local gravity are visible at the surface or hidden in the near-subsurface. Combining radar with local gravity and high-resolution imaging will break this ambiguity, separating signatures of the primordial interior from those of impact-induced resurfacing and providing the first comprehensive characterisation of the internal structure and post-impact evolution of a rubble-pile asteroid.

\section{Conclusions}

We used simulations to investigate how ejecta re-accumulation and different interior structures modify the post-impact surface and gravity field of Dimorphos. Our main conclusions are: 

1. The DART impact ejects $\approx13$\% of Dimorphos's mass, most of which remains gravitationally bound and re-accretes within the first few hours, producing near-global resurfacing. 

2. The re-accumulated ejecta blanket is strongly asymmetric, locally exceeding 8~m around the depression rim but reaching only a few tens of centimetres near the antipode. Large boulders may locally shield the underlying terrain and remain exposed above the blanket.

3. The large-scale deposition pattern is similar for both interior models. The central boulder mainly affects the local emplacement of boulders.

4. The impact modifies the surface gravity field through excavation, ejecta deposition, and redistribution of metre-scale boulders, producing changes up to a few $10^{-3}$~mm~s$^{-2}$, or $\approx$10\% of the pre-impact surface gravity in some regions. The central boulder increases the absolute gravity and produces a Bouguer anomaly of similar magnitude within the impact depression.

These post-impact modifications will need to be considered when combining Hera surface, gravity, and radar measurements to constrain the internal structure of Dimorphos.

\begin{acknowledgements}
      This research was funded by the Research Foundation - Flanders (FWO), grant number (12A5R26N).
\end{acknowledgements}
\vspace{-0.5cm}


\bibliographystyle{aa}  
\bibliography{refs.bib}

\newpage

\appendix

\section{Numerical method}
\setcounter{figure}{0}

\renewcommand{\thefigure}{A.\arabic{figure}}

\noindent The Bern SPH code was initially developed to model the collisional fragmentation of rocky bodies \citep{Benz1999}, later parallelised and further extended to model porous and granular materials. The current implementation includes a tensile fracture model \citep{Benz1999}, a $P-\alpha$-based porosity model \citep{Jutzi2008}, pressure-dependent strength formulations \citep{Collins2004}, and self-gravity. Bern SPH has been validated and benchmarked in a number of studies (e.g. \citealt{Luther2022, Ormo2022}).

The impact configuration, target properties, and treatment of the binary environment used in this study are described in Sect.~\ref{sec:methods}. Here, we provide additional details of the porosity model and of the numerical treatment used to follow the late-stage evolution of the ejecta.

\subsection{Porosity model}

Porosity is treated using the $P$--$\alpha$ compaction model \citet{Jutzi2008}, in which the distension, $\alpha = {\rho_{\rm s}}/{\rho}$, relates the density of the fully compacted solid, $\rho_{\rm s}$, to the bulk density, $\rho$. Compression progressively reduces the distension from its initial value, $\alpha_0$, towards unity as pore space is removed. We adopted a single power-law crush curve between the elastic pressure, $P_{\rm e}$, and the pressure for complete compaction, $P_{\rm s}$, via
\begin{equation}
\alpha(P) =
\begin{cases}
\alpha_0, & P \leq P_{\rm e},\\[2mm]
1+(\alpha_0-1)
\left(\dfrac{P_{\rm s}-P}{P_{\rm s}-P_{\rm e}}\right)^n,
& P_{\rm e}<P<P_{\rm s},\\[3mm]
1, & P\geq P_{\rm s},
\end{cases}
\end{equation}
where $n$ controls the shape of the crush curve. The adopted compaction parameters are summarised in Table~\ref{table:model_parameters}.

\subsection{Late-stage evolution}
\label{app:late_stage}

Following the evolution of a DART-scale impact for several hours is computationally demanding because the initial shock phase and the subsequent ejecta re-accumulation occur on very different timescales. Thus, we used the late-stage evolution approach of \citet{Raducan2022,Jutzi2022}. Once the shock has propagated through the target and the evolution is dominated by low-velocity granular motion and self-gravity, the effective elastic moduli are progressively reduced, allowing substantially larger numerical time steps.

At $t=5$, 50, and 500~s after impact, the bulk modulus is reduced to $A=359$~MPa, $3.59$~MPa, and $35.9$~kPa, respectively, with the shear modulus reduced proportionally. This procedure makes it possible to follow the ejecta evolution over several hours, while preserving the gravity-dominated dynamics relevant to re-accumulation. The approach has previously been benchmarked and tested against laboratory impact experiments \citep{Luther2022,Ormo2022}.

\begin{table}[t!]
    \footnotesize
        \caption{Target material model parameters for our impact simulations.}
        \begin{tabular}{lll}
    \bf{Description}        & \bf{Boulders} & \bf{Target matrix}  \\
    \hline
    Material             & SiO$_2$ & SiO$_2$ \\
        Equation of state (EOS)    & Tillotson$^{a,b}$ & Tillotson$^{a,b}$ \\
        Yield criterion             & Tensile$^c$ & Drucker-Prager \\
        Initial bulk modulus, A (GPa) & 35.9 & 35.9 \\   
        \hline
        \bf{Drucker-Prager criterion}\\
        Cohesion, $Y$ (Pa)   & 10$^7$ & 0.1 \\
        Limiting strength, $Y_{m}$ (GPa)  & 1.0 & 0.1  \\
        Internal friction coefficient, $f$    & 0.8 & 0.6 \\
        \hline
        \bf{Porosity model ($P-\alpha$)$^d$}           \\ 
        Grain density, $\rho_g$ (g~cm$^{-3}$)  & 3.2 & 3.2 \\             
        Initial porosity, $\phi_0$              & 10\% & 45\% \\
        Initial distension, $\alpha_0$          & 1.11 & 1.82  \\
        Solid pressure, $P_s$ (MPa)             & 40 & 20  \\
        Elastic pressure, $P_e$ (MPa)           & 1 & 1 \\
        Exponent, $n$                           & 2 & 2 \\
        \hline
    \multicolumn{3}{l}{
    $^a$\cite{Tillotson1962};
    $^b$\cite{Melosh2007};
    $^c$\cite{Jutzi2015};
    $^d$\cite{Jutzi2008}.}
        \end{tabular}
        \label{table:model_parameters}
\end{table}

\subsection{Target assembly}
\label{app:boulders}

To obtain a realistic configuration of angular, interlocking boulders, we generated the initial rubble-pile structure as a gravitational aggregate of non-spherical particles using GRAINS \citep{Ferrari2022}. The boulders were represented as polyhedra, providing a more realistic description of asteroid blocks than spherical particles and allowing mechanical interlocking between neighbouring boulders. The relaxed aggregate was then converted to SPH, preserving the individual boulder shapes and positions. Boulders smaller than 2.2~m were below the numerical resolution and were therefore not represented explicitly; together with finer granular material, they were incorporated into the matrix material filling the void space between the resolved boulders.
The resulting aggregate was trimmed to reproduce the adopted pre-impact dimensions of Dimorphos \citep{Daly2023}. For the central-boulder model, an additional 64~m boulder was placed at the centre of the rubble-pile.

\subsection{Ejecta re-accumulation timescales}
\label{app:re-accumulation}

   \begin{figure}[h!]
   \centering
   \includegraphics[width=\linewidth]{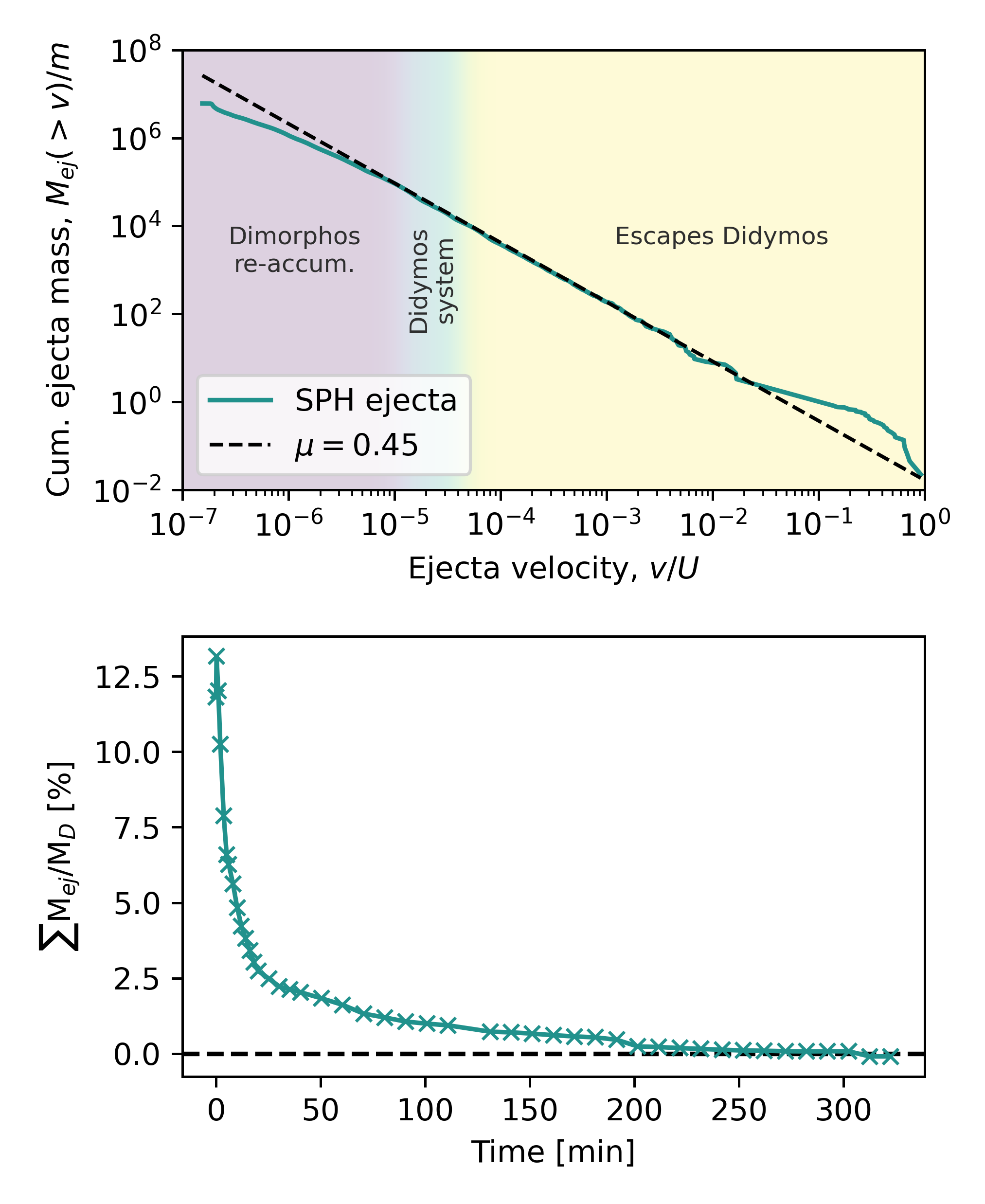}
      \caption{Ejecta mass distribution and re-accretion timescales following the DART impact.
      Top: Cumulative ejecta mass as a function of normalised ejection velocity ($v/U$), where $U$ is the DART impact velocity. The dashed line shows the best-fitting power-law with exponent $\mu = 0.45$. The shaded regions indicate ejecta that re-accretes onto Dimorphos, remains bound to the Didymos system, or escapes the binary.
      Bottom: Fraction of the ejected mass remaining in orbit over time. Most re-accretion occurs within the first 30 min after impact. After 6 h, 94\% of the ejecta has re-accreted or escaped, while the remaining $\sim$6\% is slowly orbiting material that may require days to weeks to re-accrete. }
         \label{mass_time}
   \end{figure}

The DART impact ejects approximately 13\% of the total mass of Dimorphos (Fig.~\ref{mass_time}). The cumulative ejecta mass--velocity distribution follows the expected power-law behaviour over several orders of magnitude in velocity, with an exponent of $\mu = 0.45$, consistent with other studies \citep{Raducan2019, Raducan2022, Cheng2024}. The majority of the ejecta are launched at very low velocities, with most material travelling at only a few centimetres per second (Fig.~\ref{mass_time}).

The low ejection velocities leave most of the excavated material gravitationally bound to the Didymos system. $\approx$12\% of the mass of Dimorphos eventually re-accumulates onto the surface of Dimorphos, while $\approx$1\% is either captured by Didymos or remains gravitationally bound within the binary system. Only $\approx$0.2\% of the mass reaches velocities exceeding the escape velocity of the Didymos system. 

The resulting momentum transfer corresponds to a heliocentric momentum enhancement factor of $\beta_{\rm orb} = 1.95$, defined as the ratio of the total momentum transferred to the Didymos system, including the contribution from ejecta escaping the binary, to the incident momentum of the impactor \citep{Makadia2026}. Considering instead only the momentum carried away by ejecta escaping from Dimorphos itself, we obtained a Dimorphos-scale momentum enhancement factor of $\beta \approx 3.0$.

The temporal evolution of the ejecta re-accretion is shown in the lower panel of Fig.~\ref{mass_time}. Re-accretion occurs rapidly during the first tens of minutes following the impact. Approximately 80\% of the material that returns to Dimorphos is re-accreted within the first 30 min after impact. Thereafter, the re-accretion rate decreases substantially as only the slowest ejecta remain in orbit around the binary. After approximately 6 h, approximately 94\% of the total ejected mass has either re-accreted onto one of the two bodies or escaped the system.

Because of the long dynamical timescales associated with the slowest ejecta, we do not continue tracking the remaining $\sim$6\% of the ejecta beyond the end of the simulation. This residual material consists of particles on long-lived bound trajectories that may require days to weeks to either re-impact one of the bodies or escape following subsequent gravitational perturbations. As this represents only a small fraction of the total ejecta mass, neglecting its re-accumulation has a negligible effect on the gravity field and surface evolution considered in this study.

\subsection{Ejecta layer calculation}
\label{app:layer}

The ejecta blanket was computed from the final SPH particle distribution at $t=6$ h after impact, considering only particles that had re-accreted onto Dimorphos. Surface particles were identified within an ellipsoidal shell surrounding the final body shape, while particles with velocities exceeding 5 cm s$^{-1}$ were excluded as they had not yet settled.

The surface was discretised into $1^\circ \times 1^\circ$ longitude--latitude bins. For each particle, the deposited volume was calculated as $V_i=m_i/\rho_{{\rm bulk},i}$, where $\rho_{{\rm bulk},i}$ is the bulk density of the corresponding matrix or boulder material. Because re-accreted matrix material is expected to form a more porous deposit, but the associated pore space is unresolved by the SPH simulations, we assumed a post-impact matrix bulk density equal to 80\% of its pre-impact value. Boulder particles retain their original bulk density. This assumption affects the absolute layer thickness but not its spatial distribution; if the deposited matrix is more porous, the reported thicknesses should therefore be regarded as conservative estimates. The deposited volume in each cell was calculated as $V_{\rm cell}=\sum_i V_i$ and the equivalent ejecta layer thickness was calculated as $h=V_{\rm cell}/A_{\rm cell}$, where $A_{\rm cell}$ is the cell surface area.

\subsection{Gravity calculation}
\label{app:gravity}

Effective surface-gravity maps were reconstructed directly from the relaxed particle distribution of the Bern SPH simulations. The calculation includes the self-gravity of Dimorphos, the gravitational attraction of Didymos, while the centrifugal acceleration in the co-rotating binary reference frame.

The post-impact calculations include only particles that had settled on Dimorphos by $t=6$~h. The particle positions were transformed into a reference frame centred on the mass-weighted centre of mass of Dimorphos. The surface was defined on a regular longitude--latitude grid at $1^{\circ} \times 1^{\circ}$ resolution. Within each bin, the particle with the largest radial distance from the centre of mass was used to define the local surface radius. The self-gravitational acceleration of Dimorphos at each surface point, $\boldsymbol{r}_i$, was calculated by direct summation over the SPH particles.
Didymos was represented as a point mass ($M_{\rm Didy} = 5.28\times10^{11}$~kg) located at a centre-to-centre distance of 1.206~km from Dimorphos \citep{Cheng2023}. Although this approximation neglects the higher order gravitational harmonics associated with the irregular shape of Didymos, it accurately reproduces the dominant gravitational perturbation on Dimorphos and is sufficient for the purposes of this study. The gravitational acceleration due to Didymos was calculated at each surface point as
\begin{equation}
\boldsymbol{g}_{\rm Didy}(\boldsymbol{r}_i)
= G M_{\rm Didy}
\frac{\boldsymbol{r}_{\rm Didy}-\boldsymbol{r}_i}
{\left|\boldsymbol{r}_{\rm Didy}-\boldsymbol{r}_i\right|^3}.
\label{eq}
\end{equation}
The tidal acceleration was defined as the differential gravitational acceleration due to Didymos relative to its value at the centre of mass of Dimorphos,
\begin{equation}
\boldsymbol{g}_{\rm tidal}(\boldsymbol{r}_i)
=
\boldsymbol{g}_{\rm Didy}(\boldsymbol{r}_i)
-
\boldsymbol{g}_{\rm Didy}(\boldsymbol{0}).
\end{equation}
The centrifugal acceleration in the co-rotating frame is
\begin{equation}
\boldsymbol{g}_{\rm cf}(\boldsymbol{r}_i)
=
\omega^2
\left[
\boldsymbol{s}(\boldsymbol{r}_i)-\boldsymbol{s}(\boldsymbol{0})
\right],
\end{equation}
where $\omega = \left[{G\left(M_{\rm Didy}+M_{\rm D}\right)}/{a^3}\right]^{1/2}$ is the orbital angular frequency, and $\boldsymbol{s}$ is the position relative to the rotation axis (which passes through the system barycentre), projected into the orbital $x-y$ plane. The total effective surface gravity is
\begin{equation}
\boldsymbol{g}_{\rm tot} = \boldsymbol{g}_{\rm self} + \boldsymbol{g}_{\rm tidal} + \boldsymbol{g}_{\rm cf}.
\end{equation}

\begin{figure*}[h!]
   \centering
   \includegraphics[width=\linewidth]{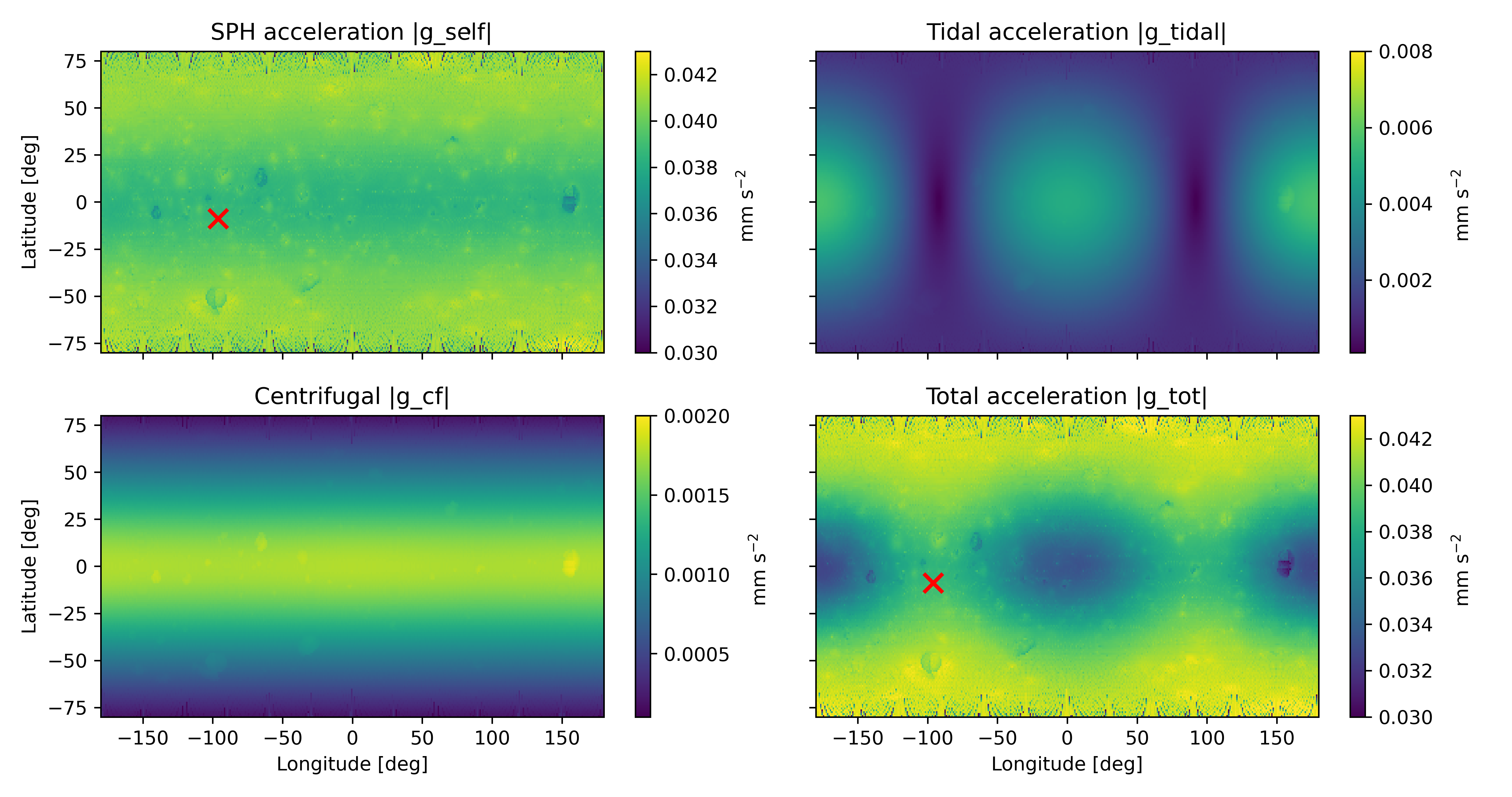}
   \caption{Effective surface-gravity maps of Dimorphos for the homogeneous rubble-pile case, pre-impact, in longitude–latitude projection. The four panels show the SPH self-gravity, the tidal acceleration, the centrifugal acceleration, and the total effective surface gravity (in mm s$^{-2}$). The red cross marks the DART impact location.}
    \label{gravity_stl_initial}
    \end{figure*}

\begin{figure*}[h!]
   \centering
   \includegraphics[width=\linewidth]{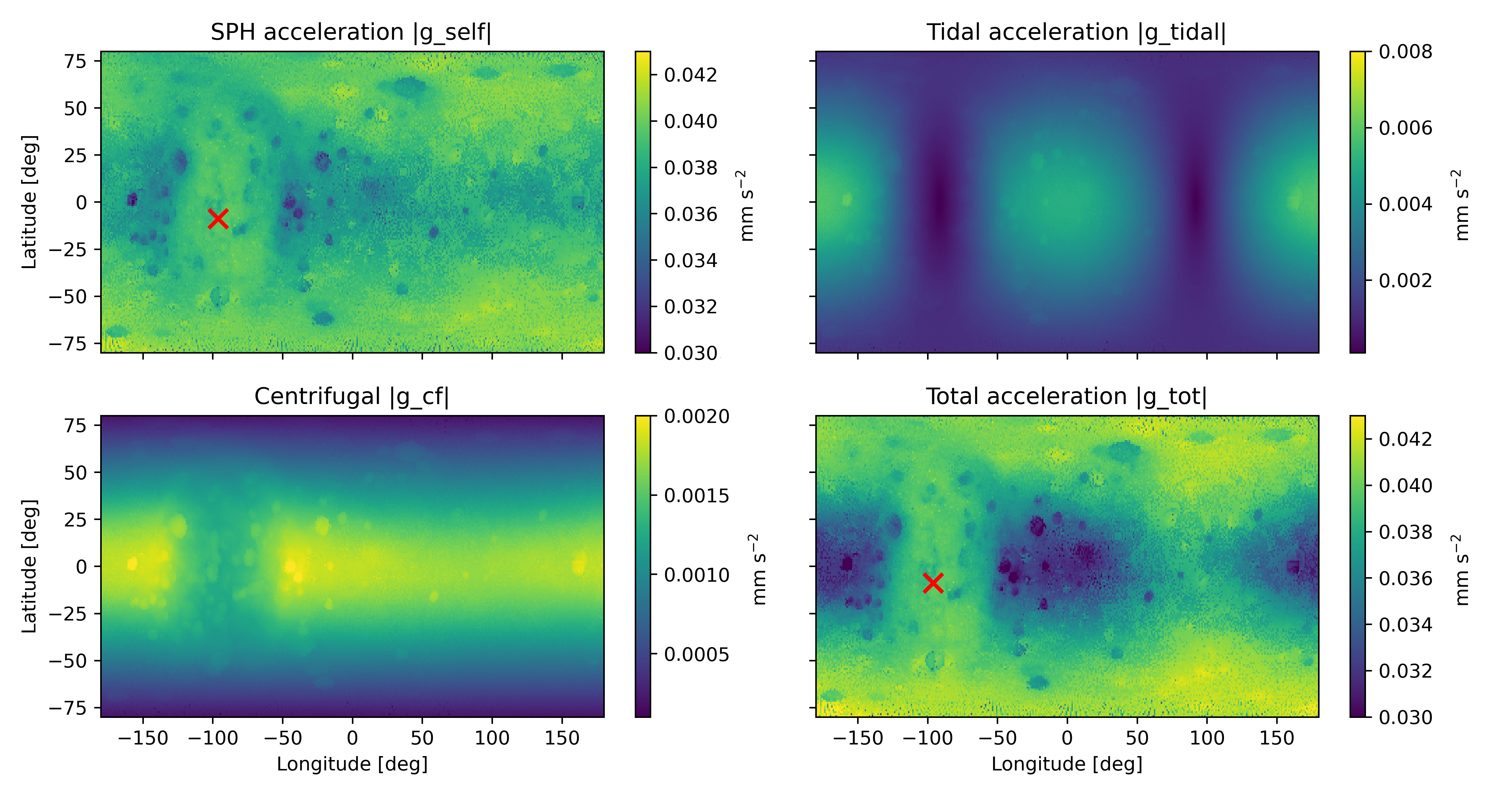}
   \caption{Same as Fig. \ref{gravity_stl_initial} but post-impact.}
    \label{gravity_stl_final}%
    \end{figure*}

\begin{figure*}[h!]
   \centering
   \includegraphics[width=\linewidth]{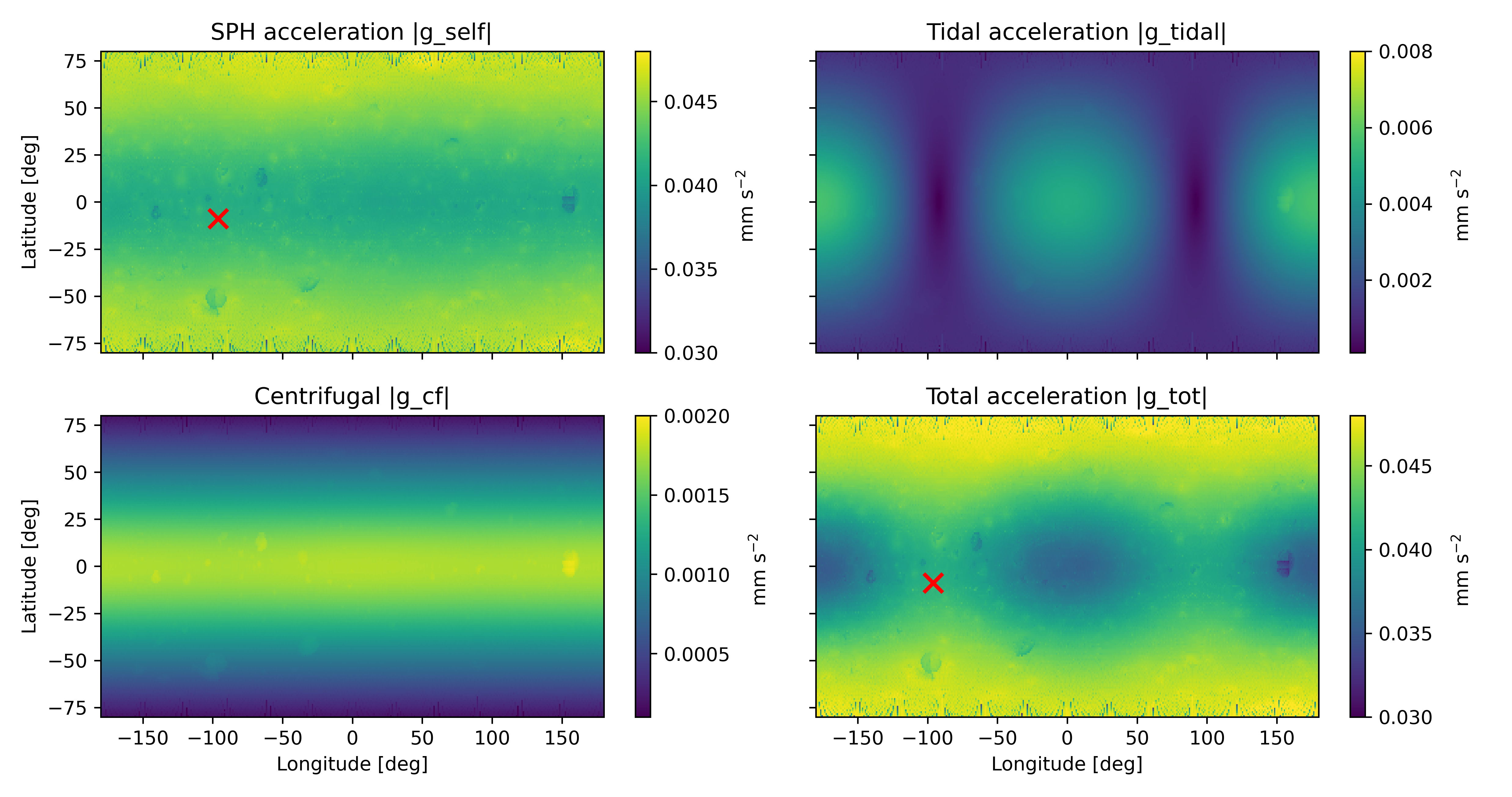}
   \caption{Effective surface-gravity maps of Dimorphos for the rubble-pile with a central boulder case, pre-impact, in longitude–latitude projection. The four panels show the SPH self-gravity, the tidal acceleration, the centrifugal acceleration, and the total effective surface gravity (in mm s$^{-2}$). The red cross marks the DART impact location.}
    \label{gravity_cb_initial}
    \end{figure*}

\begin{figure*}[h!]
   \centering
   \includegraphics[width=\linewidth]{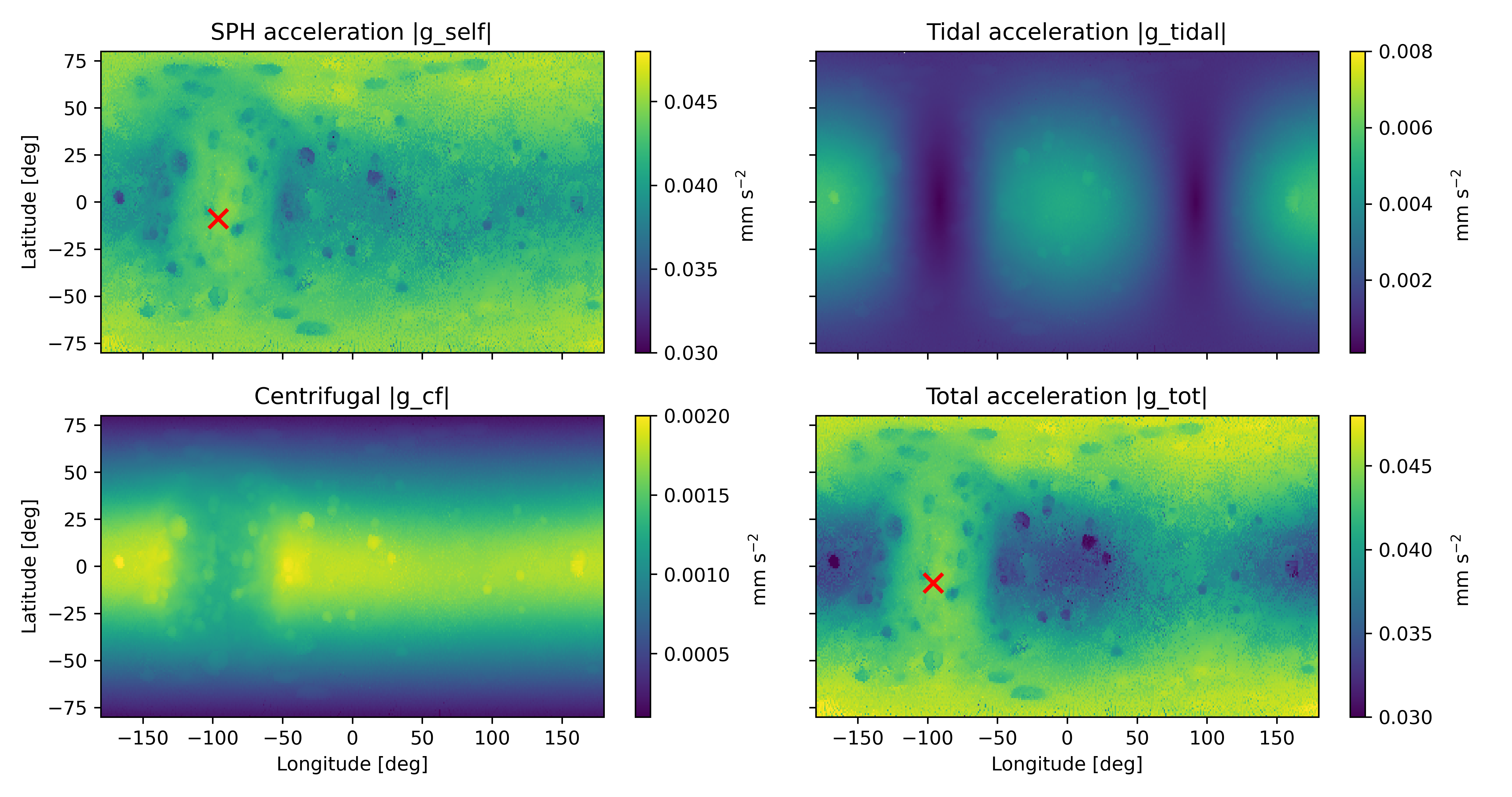}
   \caption{Same as Fig. \ref{gravity_cb_initial} but post-impact.}
    \label{gravity_cb_final}
    \end{figure*}

The Cartesian components and magnitudes of the self-gravitational, tidal, centrifugal, and total accelerations were recorded at each surface point. For visualisation, the acceleration magnitudes were mapped onto a $1^{\circ}\times 1^{\circ}$ longitude--latitude grid, using the median value of the surface points within each bin.
Figures~\ref{gravity_stl_initial}--\ref{gravity_cb_final} show the effective surface-gravity maps for the two interior structures, before and after the DART impact.

\clearpage

\clearpage

\twocolumn[{
  \centering
  \includegraphics[width=0.85\textwidth]{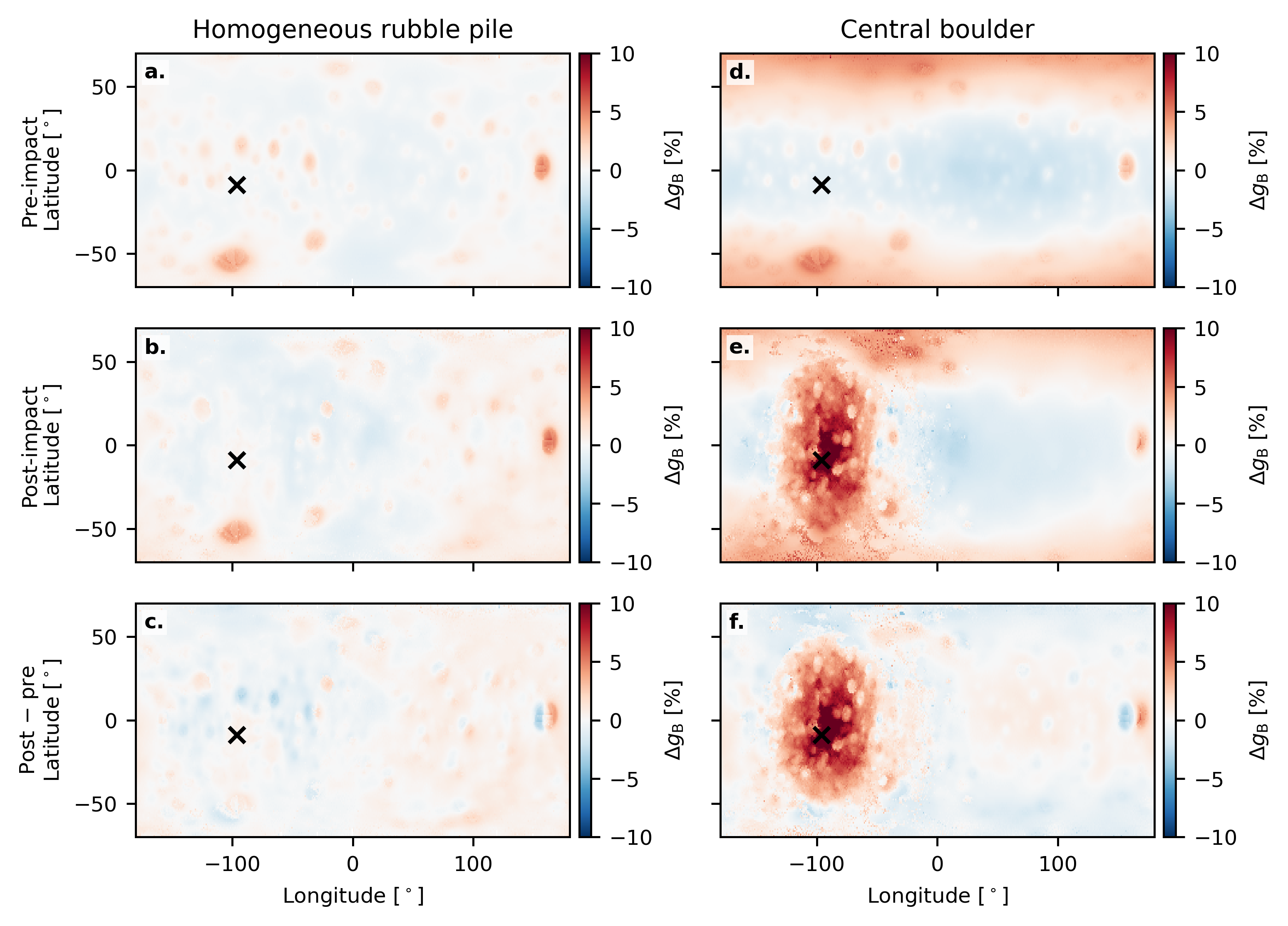}
  \captionof{figure}{Normalised Bouguer anomaly, $\Delta g_{\rm B}$, for the homogeneous rubble-pile (a--c) and central-boulder (d--f) models. Rows show the anomaly before the impact (a, d), 6~h after the impact (b, e), and the difference between the two (c, f). The anomaly is the difference between the self-gravity of the simulated body and that of a uniform-density body of identical shape and mass, normalised by the latter. The black cross marks the DART impact location. Only the central-boulder model develops a positive anomaly within the impact depression, produced by the reduced distance between the surface and the dense central boulder.}
  \label{fig:bouguer}
  \vspace{1.5em}
}]
\subsection{Bouguer gravity anomaly}\label{app:bouguer}
To isolate the contribution of internal density heterogeneities from that of the body shape, we computed a normalised Bouguer anomaly at each surface point. For each particle distribution, we constructed a reference body with the same particle positions (and shape), but uniform density. Each particle was assigned the mass $m_i^{\rm u} = \bar{\rho}\,V_i$, where $V_i = m_i/\rho_i$ is the particle volume and $\bar{\rho} = M_{\rm D}/\sum_i V_i$, so that the total mass is conserved.

The self-gravitational accelerations of the real and reference bodies, $\vec{g}_{\rm self}$ and $\vec{g}^{\,\rm u}_{\rm self}$, were computed by direct summation at the same surface points (Appendix~\ref{app:gravity}). The Bouguer anomaly is their difference projected along the local gravity direction and normalised by the reference value,
\begin{equation}
  \Delta g_{\rm B} = \frac{\left(\vec{g}_{\rm self} -
  \vec{g}^{\,\rm u}_{\rm self}\right)\cdot\hat{\vec{g}}_{\rm self}}
  {\left|\vec{g}^{\,\rm u}_{\rm self}\right|}.
\end{equation}

The tidal and centrifugal accelerations are the same for both bodies at a given point and, therefore, they cancel out. Positive values correspond to locally denser-than-average material. Figure~\ref{fig:bouguer} shows the change in surface radius together with the Bouguer anomaly before and after the impact. The homogeneous rubble-pile model shows only small anomalies associated with individual boulders. 

In the central-boulder model, the dense interior boulder produces a weak pole-to-equator pattern prior to impact because its contribution to the surface gravity varies with position over the non-spherical body. After impact, excavation lowers the local surface by $\approx$20~m, reducing the distance between the surface and the dense interior boulder. This enhances the boulder’s contribution to the local gravity field and produces a Bouguer anomaly of $\approx$10\%, which is absent from the homogeneous model.

\end{document}